\documentclass[trackchanges]{aastex701}

\begin{document}

\title{How sky subtraction scales affect Intracluster light measurements}

\author[orcid=0009-0001-2377-272X,sname='Golini']{Giulia Golini}
\affiliation{School of Physics, University of New South Wales, NSW 2052, Australia}
\email[show]{g.golini@unsw.edu.au}

\author[orcid=0000-0002-9796-1363, sname='Brough']{Sarah Brough} 
\affiliation{School of Physics, University of New South Wales, NSW 2052, Australia}
\email{s.brough@unsw.edu.au}

\author[orcid=0000-0001-7847-0393, sname='Montes']{Mireia Montes} 
\affiliation{Institute of Space Sciences (ICE, CSIC), Campus UAB, Carrer de Can Magrans, s/n, 08193 Barcelona, Spain.}
\email{mmontes@ice.csic.es}

\author[orcid=0000-0003-4859-3290, sname='Watkins']{Aaron Watkins} 
\affiliation{Centre for Astrophysics Research, University of Hertfordshire, College Lane, Hatfield, AL10 9AB, UK}
\email{a.watkins@herts.ac.uk}

\begin{abstract}

Deep imaging surveys are opening a new window on the low surface brightness universe. However, the recovery of diffuse emission depends on the survey depth and how the sky background is treated. Estimating the sky on scales comparable to astrophysical sources risks biasing flux measurements.
This effect is particularly significant for intracluster light, the most extended diffuse baryonic component of galaxy clusters. We show in a $M_{500}$--$z$ plane that the sky subtraction scales employed in the Legacy Survey of Space and Time Data Preview 2 and Hyper Suprime-Cam Public Data Release 3 (HSC-PDR3) impact clusters with $M_{500}\gtrsim10^{13.4}\,\mathrm{M_\odot}$ at all redshifts, whereas HSC-PDR2 only oversubtracts the most massive systems at $z < 0.1$.
The larger sky subtraction scale anticipated for the full LSST survey will mitigate this bias. 

\end{abstract}

\section{Rationale} 
\label{sect:rationale} 

Modern survey astronomy is no longer limited by data volume, but by how effectively those data are processed. Wide-field surveys such as the European Space Agency’s Euclid mission \citep{laureijs2011euclid}
and the Vera C. Rubin Observatory's Legacy Survey of Space and Time (LSST; \citealp{ivezic2019lsst}) provide unprecedented imaging datasets, yet their scientific potential for low surface brightness (LSB) studies depends critically on the treatment of diffuse emission.  
To achieve the required photometric precision, their pipelines are optimized for compact sources and cosmological analyses \citep[see e.g.][]{2026romelli,RTN-115}, often relying on local background subtraction to remove instrumental and atmospheric variations. However, if extended low surface brightness emission is not properly
masked, it can be mistaken for sky and subtracted, leading to sky oversubtraction \citep[e.g.,][]{kelvin2023sky}. This effect can produce artificial 'dark holes’ around bright galaxies and underestimate the
brightness of their outskirts, though dedicated detection tools can mitigate the bias. Accurate background modeling and masking remains a key challenge, requiring large-scale background removal without compromising extended LSB structures.

This issue is particularly important for studies of intracluster light \citep[ICL; e.g.,][]{mihos2015intragroup,montes2022faint}. As the most extended single baryonic component in galaxy clusters, the ICL is vulnerable to oversubtraction, yet its analysis is key, as it traces cluster assembly and dynamical history. The challenge of sky subtraction for LSB studies motivated the development of dedicated processing strategies aimed at preserving diffuse emission 
\citep[see e.g., ][]{duc2015atlas3d,montes2022new, kelvin2023sky,golini2024ultra, zaritsky2024lights, watkins2024strategies}. Nevertheless, standard survey products remain optimized for their primary cosmological science goals, meaning that diffuse structures on large scales may still be underestimated. In the context of ICL studies, this raises a practical question: over what range of halo mass and redshift can ICL be reliably recovered from standard survey imaging products? We address this question by comparing sky subtraction scales in current wide field surveys with the expected spatial extent of astronomical systems across halo mass and redshift,  identifying regions
where diffuse emission becomes increasingly susceptible to sky oversubtraction. The characteristic sky subtraction scale should be viewed as a threshold of ``increasing risk'' of oversubtraction, with true oversubtraction depending on source properties, observational conditions and image analysis.

\section{analysis and DISCUSSION} 
\label{sec:resultsdiscussion}

Our goal is to estimate the scales at which sky subtraction may bias ICL measurements. 
To relate survey sky-subtraction scales to the physical extent of galaxies and clusters, we characterize system size through the effective radius, $R_{\rm e}$, as a function of halo mass, $M_{500}$.
Assuming an NFW halo profile \citep{1997Navarrofw}, we convert $M_{500}$ to $M_{200}$, derive $R_{200}$ \citep{Hu_2003} and $R_{\rm e}\simeq0.015 R_{200}$ for the central galaxy \citep{kravtsov2013size}.

\begin{figure*}[ht!]
\includegraphics[width=\textwidth]{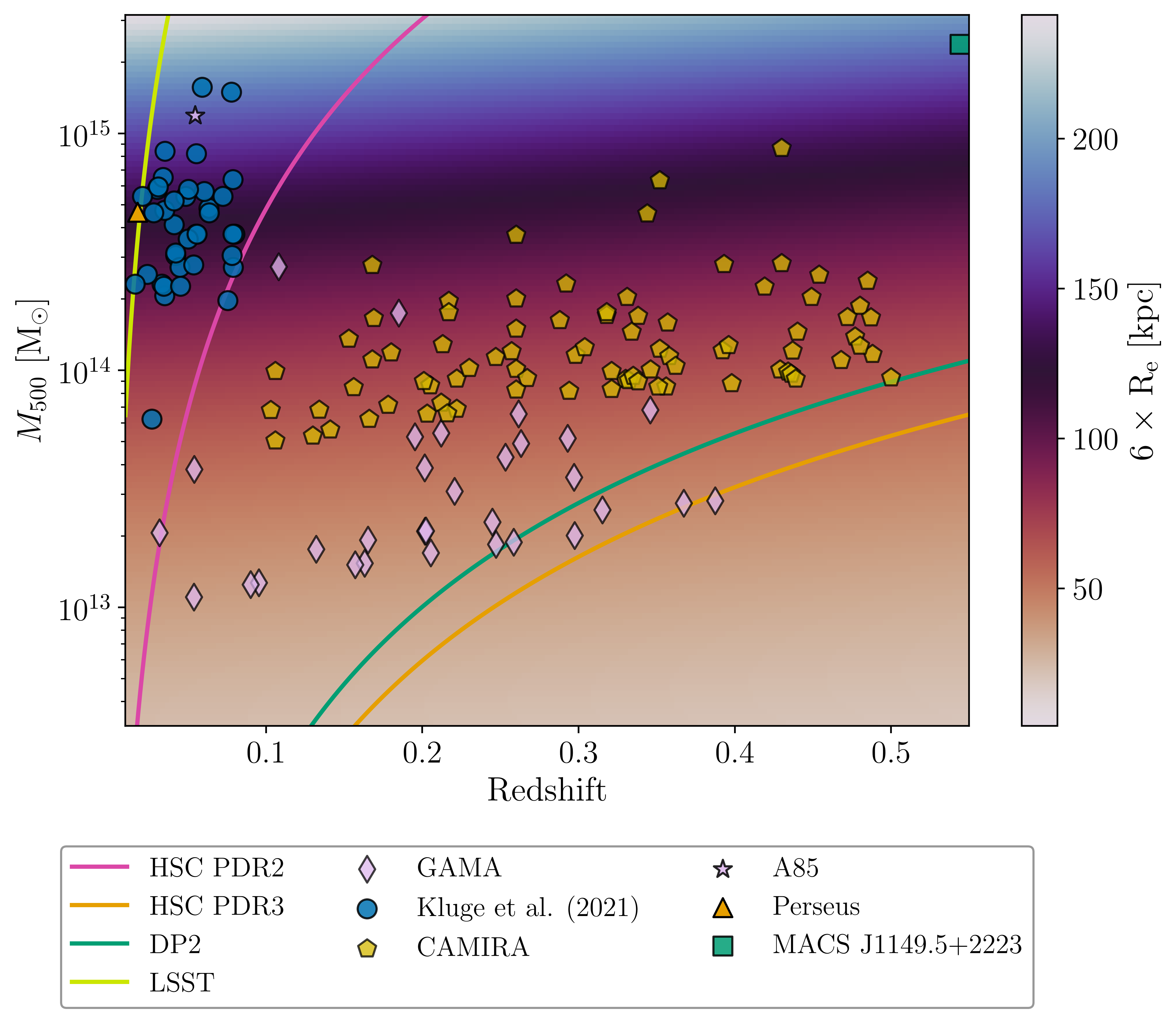}
\caption{Halo mass-redshift diagram comparing survey sky-subtraction scales with a diffuse extent of 6 $\times$ R$_{\rm e}$. Systems left of each curve are increasingly susceptible to sky oversubtraction. Literature cluster measurements are overplotted.}
\label{fig1}
\end{figure*}

We adopt $6R_{\rm e}$ as a representative ICL extent \citep{iodice2016fornax}. Figure~\ref{fig1} compares this extent with the scales corresponding to the sky estimation bin size adopted by current and future surveys. %on their final coadded data products. 
For Hyper Suprime-Cam Public Data Release 2 (HSC-PDR2), the sky background is estimated on a $1024\times1024$ pixel grid, corresponding to $\sim2.8^\prime\times2.8^\prime$ \citep{aihara2019second}. HSC-PDR3 uses a $128\times128$ pixel grid ($\sim21.5^{\prime\prime}\times21.5^{\prime\prime}$; \citealp{aihara2022third}) and LSST Data Preview 2 (DP2) adopts a $128\times128$ pixel grid ($\sim25.6^{\prime\prime}\times25.6^{\prime\prime}$). %with sixth-order spline interpolation \citep{RTN-115}.
For the full ten-year LSST survey, the sky modeling bin size is expected to be $4096 \times 4096$ pixel ($\sim13.6^\prime \times 13.6^\prime$; A. Watkins priv. communication).
For completeness, Euclid uses a $256\times256$-pixel ($25.6''\times25.6''$) background grid \citep{2026romelli}, although LSB-optimized reductions have been developed to mitigate sky oversubtraction (e.g., \citealp{euclidcollaboration2026euclidquickdatarelease}; P.~Dimauro et al. (
in preparation)).
The curves in Fig.~\ref{fig1} mark the sky-subtraction scales of the different surveys. Systems to the left of a curve are larger than the adopted sky scale and are therefore susceptible to oversubtraction.
For reference, we overplot halo masses derived from the CAMIRA cluster catalog \citep[][]{oguri2018optically}, the Galaxy and Mass Assembly (GAMA) Galaxy Group
Catalogue \citep{Robotham2011}, the \cite{kluge2021photometric} cluster sample, Abell 85 \citep[][]{montes2021buildup}, Perseus \citep[][]{kluge2025euclid}, and MACSJ1149.5$+$2223 \citep{gonzalez2021discovery}. HSC-PDR3 and LSST DP2 may oversubtract ICL in groups and clusters at all redshifts, biasing its measurement. HSC-PDR2's scale remains effective over a larger region of the sky, only affecting the most nearby and massive clusters ($z<0.1$; $M_{500}\gtrsim10^{14.1}\,\mathrm{M_\odot}$). The ten-year LSST survey offers the broadest parameter space for ICL studies, affecting only the nearest, most extended systems. Varying the adopted ICL extent (6 $\times$ R$_{\rm e}$) by a factor of two shifts the curves but does not alter the qualitative conclusion that low-redshift massive clusters are most susceptible to oversubtraction.

Survey processing can be as important as depth for LSB science. Fig. \ref{fig1} reveals that sky subtraction preferentially affects nearby massive systems, providing a simple framework for evaluating data suitability for ICL studies.

%% For this sample we use BibTeX plus aasjournalv7.bst to generate the

\bibliography{sample701}{}
\bibliographystyle{aasjournalv7}



\end{document}